\documentclass[12pt]{article}

\input epsf

\begin{document}

\title{\bf The Googly Amplitudes in Gauge Theory}

\author{Chuan-Jie Zhu\thanks{Supported in
part by fund from the National Natural Science Foundation of China
with grant Number
90103004.} \\
Institute of Theoretical Physics,
Chinese Academy of Sciences\\
P. O. Box 2735,  Beijing 100080, P. R. China}

\maketitle


\begin{abstract}
The googly amplitudes in gauge theory are computed by using the
off shell MHV vertices with the  newly proposed rules of Cachazo,
Svrcek and Witten. The result is   in agreement with the
previously well-known results. In particular we also obtain a
simple result for the all negative but one positive helicity
amplitude when one of the external line is off shell.
\end{abstract}


\section{Introduction}

In a recent paper \cite{Wittena} the calculation of the
perturbative amplitudes in gauge theory was  reformulated by using
the off shell MHV vertices. These   MHV vertices are the usual
tree level MHV scattering amplitudes in gauge theory \cite{Parkea,
Giele}, continued off shell in a particular fashion as given in
\cite{Wittena}. Some sample calculations were done in
\cite{Wittena}, sometimes with the help of symbolic manipulation.
The correctness of the    rules was partially verified by
reproducing the known results for small number of gluons
\cite{Parkeb}.

In view of the deep connection of the gauge theory with twistor
space and string theory \cite{Penrosea,Wittenb}, one would like to
push the computation of the gauge theory amplitudes to its limit.
By doing these calculations in the new formalism, we hope to learn
more about the techniques of doing perturbative calculations in
gauge theory or QCD \cite{Berna, Dixon, Bernb}, and to discover
the inner structure of the amplitude as used in calculating loop
amplitudes \cite{Bernb, Others}. A natural question is how to
reproduce the exceptionally simple amplitudes with two positive
helicity gluons and an arbitrary number of negative helicity ones,
called googly amplitudes in \cite{Wittena}. These amplitudes were
calculated from the string theory in \cite{itp}. As we will shown
in this paper, the googly amplitudes can be calculated rather
simply by following the new approach of \cite{Wittena}. By
reproducing the previously well known results, our calculation
gives a quite strong support to the Cachao-Svrcek-Witten proposal.

This paper is organized as follows. In section 2 we review briefly
the spinor formalism and the rules of calculating the  gauge
theory amplitudes as proposed in \cite{Wittena}. For our purpose
we will prove a general result for the all negative but one
positive helicity amplitude when one of the external line is off
shell. Two general formulas with  spinors will also be proved. In
section 3 we will compute the googly amplitude in the special case
when the two positive helicity gluons are adjacent. By using an
identity we reproduce the known result for this amplitude. The
proof of the required identity is given in section 4. The
calculation of the more general googly amplitudes is only briefly
described. The detail  for their calculations and the proof of a
more general identity will appear in a separate publication.

\section{The MHV vertices and some formulas with spinors}

First let us recall the rules for calculating tree level gauge
theory amplitudes as proposed in \cite{Wittena}. We will use the
convention that all momenta are outgoing. By MHV we always mean an
amplitude with precisely two gluons of negative helicity.  If the
two gluons of negative helicity are labeled as $r,s$ (which may be any
integers from $1$ to $n$), the MHV vertices (or amplitudes) are given
as follows:
\begin{equation}
V_n =  {\langle\lambda_r,\lambda_s\rangle^4\over
\prod_{i=1}^n\langle\lambda_i, \lambda_{i+1}\rangle } .
\label{eqone}
\end{equation}
For an on shell (massless) gluon, the momentum in bispinor basis
is given as:
\begin{equation}
p_{a\dot a} = \sigma^\mu_{a\dot a} p_\mu =
\lambda_a \tilde{ \lambda}_{\dot a}.
\end{equation}
For an off shell momentum, we can no longer define $\lambda_a$ as
above. The off-shell continuation given in \cite{Wittena} is to
choose an arbitrary spinor $\tilde\eta^{\dot a}$ and then to
define $\lambda_a$ as follows:
\begin{equation}
\lambda_a = p_{a\dot a}\tilde{\eta}^{\dot a}.
\end{equation}
For an on shell momentum $p$, we will use the notation
$\lambda_{pa}$ which is proportional to $\lambda_a$:
\begin{equation}
\lambda_{pa} \equiv  p_{a\dot a} \tilde{\eta}^{\dot a} = \lambda_a
\tilde{\lambda}_{\dot a} \tilde{\eta}^{\dot a} \equiv \lambda_a
\phi_p.
\end{equation}
As demonstrated in \cite{Wittena}, it is consistent to use the same
$\tilde\eta$ for all the off shell lines (or momenta).

 \begin{figure}[ht]
    \epsfxsize=80mm%
    \hfill\epsfbox{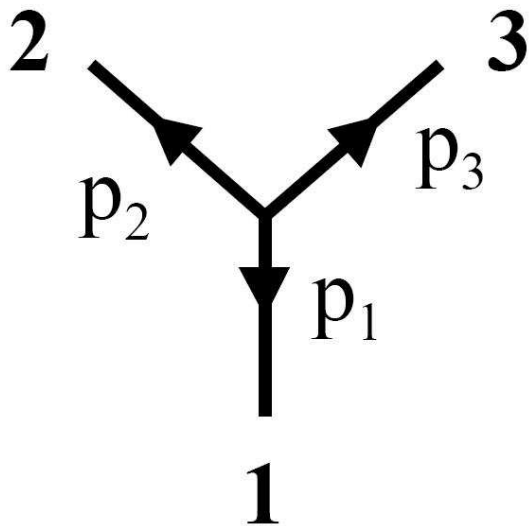}\hfill~\\
    \caption{The 3 gluon vertex. It is not important which
    gluon has positive helicity.}
    \label{figa}
   \end{figure}

The 3 gluon vertex is very important for our purpose. As shown in
Fig.~1, even if all the momenta are off shell, one can simplify
this (MHV) vertex by using momentum conservation. We have
\begin{eqnarray}
& & \lambda_1 +  \lambda_2 +\lambda_3 = 0 , \\
& & V_3 = {\langle\lambda_r,\lambda_s\rangle^4\over
\prod_{i=1}^3\langle\lambda_i, \lambda_{i+1}\rangle } =
\langle\lambda_1,\lambda_2\rangle =
\langle\lambda_2,\lambda_3\rangle
=\langle\lambda_3,\lambda_1\rangle . \label{eq3}
\end{eqnarray}

By using only MHV vertices, one can build a tree diagram by
connecting MHV vertices with propagators.  For the propagator of
momentum $p$, we assign a factor $1/p^2$. Any possible diagram
(involving only MHV vertices) will contribute to the amplitude. As
proved in \cite{Wittena}, a tree level amplitude with $q$ external
gluons of negative helicity must be obtained from an MHV tree
diagram with $q-1$ vertices. This counting can be generalized. As
our purpose is to compute amplitudes with  few external gluons
with positive helicity, one easily prove that a tree level
amplitude with $n_+$ external gluons of positive helicity will
have no contribution from any diagram containing an MHV vertex
with more than $n_++2$ lines (not necessarily all internal). The
reason is that there is not enough positive helicity lines because
an internal line must have a positive helicity at one end and a
negative helicity at the other end. Now we give a more precise
formulation.  If we assume that the number of the vertices with
exactly $i$ lines is $n_i$ ($n_i\ge 0$) and let $n_-$ denotes the
external gluons with negative helicity, we have:
\begin{eqnarray}
n_+ & = & \sum_{ i} n_i (i-3) +1, \\
n_- & = & \sum_{i} n_i + 1 .
\end{eqnarray}
For $n_+=1$ we must have $n_i=0$ for $i\ge 4$ and $n_3$ is given
by $n_--1$ as derived \cite{Wittena}). For $n_+=2$ any
contributing diagram will have exactly one MHV vertex with 4
lines, i.e. $n_4 = 1$. This paper will calculate only tree
amplitude with 1 or 2 external gluons with positive helicity.

Our first result is a general expression for the off shell amplitude with all
negative but one positive helicity:
\begin{equation}
V(1+,2-, \cdots, n-) = {p_1^2 \over \phi_2 \phi_n} \,
{1 \over {[}2,3][3,4]\cdots [n, n-1] }, \label{eq34}
\end{equation}
where only the first particle with momentum $p_1$ is off-shell and
has positive helicity. For $n=3$ this coincides with the off shell
MHV vertices given in (\ref{eqone}). Here we have used the following formula:
\begin{equation}
p_1^2 = (p_2 + p_3)^2 =  2 p_2\cdot p_3 = \langle 2 , 3\rangle [2,3].
\end{equation}
The other variant of (\ref{eq34}) is for the case when the off shell gluon
has negative helicity. We relabel this gluon to be the first gluon and the
amplitude is given as follows:
\begin{equation}
V_n(1-,2-, \cdots,r+,\cdots,
 n-) = {\phi_r^4 p_1^2 \over \phi_2 \phi_n} \,
{1\over [2,3][3,4]\cdots [n, n-1]} , \label{eq22}
\end{equation}
where the $r$-th gluon has positive helicity. Now we prove the
above results by the method of mathematical induction.

As we noted before, all contributing diagrams must contain only
the 3 gluon vertex. By using this result one can decompose a
multi-gluon amplitude into two sub-multi-gluon amplitude with less
gluons, as shown in Fig.~2. As one can see from this figure, a
multi-gluon amplitude with only one positive helicity can be
decomposed into two multi-gluon amplitudes with less gluons. The
two sub-multi-gluon amplitudes also have only one positive
helicity and contain fewer gluons. In this way, we could  prove
eq.~(\ref{eq34}) by mathematical induction.

 \begin{figure}[ht]
    \epsfxsize=100mm%
    \hfill\epsfbox{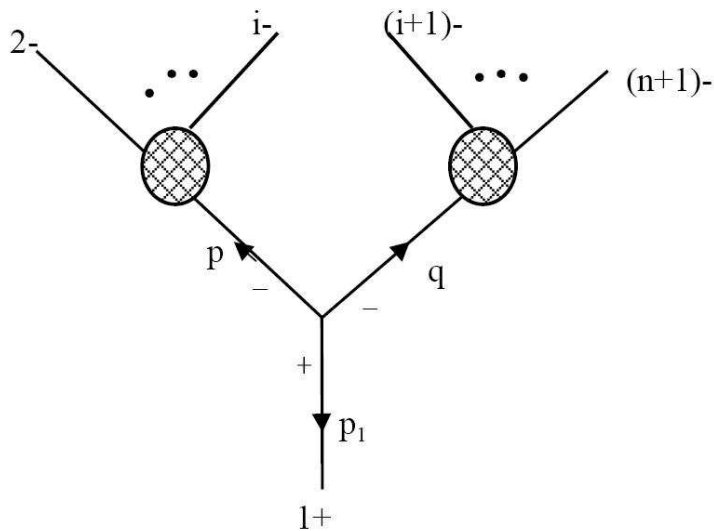}\hfill~\\
    \caption{A multi-gluon amplitude with only one positive helicity
    can be decomposed into two multi-gluon amplitudes with less gulons. The two
    sub-multi-gluon amplitudes also have only one positive helicity. A summation over
    $i$ should be understood.}
    \label{figb}
   \end{figure}

To begin the mathematical reduction, we first note that
eq.~(\ref{eq22}) is true for $n=3$. We assume that it is also true
for all $k\le n$. We will prove that is is also true for $k=n+1$.
In order to prove this, let us compute $V_{n+1}$ from Fig.~2. We
have
\begin{eqnarray}
V_{n+1} & = & \sum_{i=2}^n \left\{
{p^2 \over \phi_2 \phi_i}\, {1 \over
[2,3]\cdots [i-1,i]} \right\}  \times {1\over p^2}
\times   { \langle \lambda_p, \lambda_q \rangle^3 \over
\langle \lambda_1, \lambda_p\rangle \langle \lambda_q,
\lambda_1\rangle } \times {1\over q^2} \nonumber \\
& & \qquad
\times {q^2 \over
\phi_{i+1}\phi_{n+1}}\, {1\over [i+1,i+2]\cdots [n,n+1]} ,
\label{eqaa}
\end{eqnarray}
by using the assumed result for all less multi-gluon amplitudes. Here
\begin{eqnarray}
p = \sum_{l=2}^i  p_l , \qquad  & &
\lambda_p = \sum_{l=2}^i \lambda_l \phi_l, \\
q = \sum_{l=i+1}^{n+1}  p_l , \qquad & &
\lambda_q = \sum_{l=i+1}^{n+1} \lambda_l \phi_l .
\end{eqnarray}
The degenerate cases for $i=2$ and $i=n$ are also included
correctly in the sum in (\ref{eqaa}) as one can easily verify. It
is important to note here $\lambda_{p_2} = \lambda_2\phi_2$ which
is not identical to $\lambda_2$.

By using eq.~(\ref{eq3}), we have
\begin{equation}
V_{n+1} = {1\over \phi_2 \phi_{n+1}} \, {1\over
[2,3] \cdots [n,n+1]} \, \sum_{i=2}^n
{[i, i+1]\over \phi_i \phi_{i+1}} \, \langle \lambda_p, \lambda_q
\rangle . \label{eq13}
\end{equation}

In order to simplify the above result we prove two formulas
involving spinors. The first formula is:
\begin{equation}
{[i,j]\over \phi_i \phi_{j} } =  { [i,l]\over \phi_i \phi_{l} }
- { [j,l]\over \phi_{j} \phi_{l} } =  { [i,l]\over \phi_i \phi_{l} }
+ { [l,l]\over \phi_{j} \phi_{l} }, \label{ida}
\end{equation}
where $p_l$ is any on shell momentum (not necessarily be one of
momenta in question). To prove the above formula one can assume
$\tilde\eta^1=0$. (The general case is obtained by an $SL(2)$
transformation.) We have
\begin{eqnarray}
{  [i,j] \over \phi_i\phi_j } & =  & { \tilde\lambda_{i1}
\tilde\lambda_{j2} -\tilde\lambda_{i2}
\tilde\lambda_{j1}   \over \tilde\lambda_{i2}
\tilde\lambda_{j2} (\tilde\eta_1)^2} = { \tilde\lambda_{i1}
 \over \tilde\lambda_{i2} (\tilde\eta_1)^2}   -{ \tilde\lambda_{j1}
 \over \tilde\lambda_{j2} (\tilde\eta_1)^2}  \nonumber \\
 & = & \left[ { \tilde\lambda_{i1}
 \over \tilde\lambda_{i2} (\tilde\eta_1)^2} - { \tilde\lambda_{l1}
 \over \tilde\lambda_{l2} (\tilde\eta_1)^2} \right] +
\left[ { \tilde\lambda_{l1}
 \over \tilde\lambda_{l2} (\tilde\eta_1)^2} - { \tilde\lambda_{j1}
 \over \tilde\lambda_{j2} (\tilde\eta_1)^2} \right]  \nonumber \\
 & = & {  [i,l] \over
 \phi_i\phi_l  } + { [l,j] \over \phi_l\phi_j } .
 \end{eqnarray}

The other formula is:
\begin{equation}
B_n = \sum_{i=2}^{n-1}
{[i, i+1]\over \phi_i \phi_{i+1}} \, \langle \lambda_{p_2 + \cdots + p_i},
\lambda_{p_{i+1} + \cdots + p_n}
\rangle = (p_2 + p_3 + \cdots + p_n)^2 = p_1^2. \label{idb}
\end{equation}
where momenta conservation is used in the second equality.
Now we prove this result. By using eq.~(\ref{ida}), we have
\begin{eqnarray}
B_n & = & \sum_{i=1}^{n-1} \left[ { [i,l] \over \phi_i \phi_l } -
 {{ [}i+1,l] \over \phi_{i+1} \phi_l }  \right]  \langle \lambda_{p_2 + \cdots + p_i},
\lambda_{p_{i+1} + \cdots + p_n} \rangle \nonumber \\
 & = & \sum_{i=2}^{n} { [i,k]\over \phi_i \phi_k }\langle \lambda_{p_i},
 \lambda_{p_2 + \cdots + p_n} \rangle \nonumber \\
 & = & {1\over \phi_k} \, \sum_{i=2}^{n} [i,k] \langle \lambda_{i},
 \lambda_{p_2 + \cdots + p_n} \rangle \nonumber \\
 & = & {1\over \phi_k} \, \sum_{i=2}^{n} \tilde\lambda_{i\dot a}\tilde\lambda_k^{\dot a}
 \lambda_{ia}   \,  \lambda_{p_2 + \cdots + p_n}^a \nonumber \\
 & = & (p_2 + \cdots + p_n)_{a\dot a} (p_2 + \cdots + p_n)^{\dot b
 a} \, \tilde\eta_{\dot b} \tilde\lambda_k^{\dot a}/\phi_k \nonumber \\
 & = &  - (p_2 + \cdots + p_n)^2 \delta^{\dot b}_{\dot a}  \tilde\eta_{\dot b}
 \, \tilde\lambda_k^{\dot a}/\phi_k  =
 (p_2 + \cdots + p_n)^2.
 \end{eqnarray}
This ends the proof of eq.~(\ref{idb}).

By using eq.~(\ref{idb}) in (\ref{eq13}) we have:
\begin{equation}
V_{n+1} = {p_1^2 \over \phi_2 \phi_{n+1}} \, {1\over
[2,3] \cdots [n,n+1]} ,
\end{equation}
as announced. This finishes the proof of (\ref{eq34}) by
mathematical induction. The other case when the off shell gluon
has negative helicity is proved almost the same as above. We
needn't care much about if the positive helicity particle is in
the left blob or in the right blob in Fig.~2. The three gluon
vertex doesn't differentiate much the ordering of the two
different helicities as one can see from eq.~(\ref{eq3}).

\section{The googly amplitude}

Now we compute the $n$-particle googly amplitude. In this paper we
concentrate on the easier case when the two positive helicity
particles are adjacent. We label them  to be $1$ and $n$. As we
noted in section 2, there is only one 4 line MHV vertex in any
Feynman diagram. By using this result, the amplitude is computed
by using the diagram decomposition as shown in Fig.~3.

 \begin{figure}[ht]
    \epsfxsize=100mm%
    \hfill\epsfbox{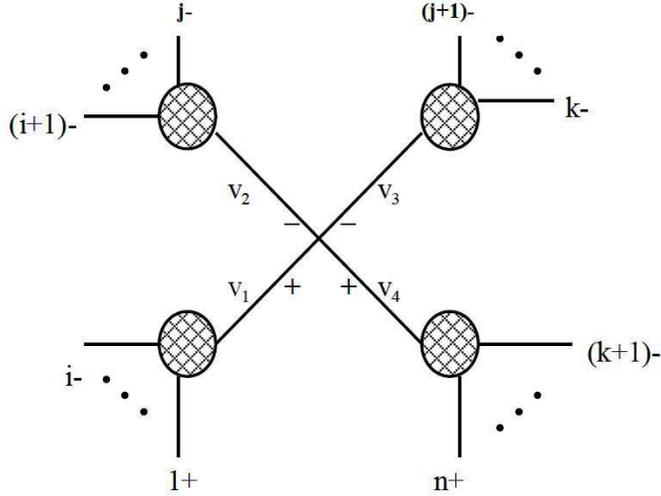}\hfill~\\
    \caption{The decomposition of the googly
    amplitude  $A_n(1+,2-,\cdots,(n-1)-,n+)$. We
    note that there is only one 4 gluon vertex.}
    \label{figa}
   \end{figure}
All the  4 blob diagrams in Fig.~3 have been computed.
By using the results proved in section 2, we have
\begin{eqnarray}
A_n & = & {\phi_1^3 \phi_n^3 \over [1,2][2,3]\cdots [n-1,n]}
\sum_{i=1}^{n-3}\sum_{j=i+1}^{n-2}\sum_{k=j+1}^{n-1}
{[i,i+1]\over \phi_i \phi_{i+1}} \,
{[j,j+1]\over \phi_j \phi_{j+1}} \,
{[k,k+1]\over \phi_k \phi_{k+1}} \, \nonumber \\
& & \qquad \times { \langle V_2, V_3 \rangle^3 \over \langle V_1,
V_2 \rangle \langle V_3, V_4 \rangle \langle V_4, V_1 \rangle } ,
\label{eqab}
\end{eqnarray}
where
\begin{eqnarray}
V_1 = \sum_{l=1}^i \lambda_l \phi_l , \qquad & &
V_2 = \sum_{l=i+1}^j \lambda_l \phi_l , \\
V_3 = \sum_{l=j+1}^k \lambda_l \phi_l , \qquad & &
V_4 = \sum_{l=k+1}^n \lambda_l \phi_l .
\end{eqnarray}

One can prove that the 3-fold summation in eq.~(\ref{eqab}) gives
exactly the required result, i.e.
\begin{eqnarray}
& & \sum_{i=1}^{n-3}\sum_{j=i+1}^{n-2}\sum_{k=j+1}^{n-1}
{[i,i+1]\over \phi_i \phi_{i+1}} \,
{[j,j+1]\over \phi_j \phi_{j+1}} \,
{[k,k+1]\over \phi_k \phi_{k+1}} \,  \nonumber \\
& & \qquad \qquad \times
{ \langle V_2, V_3 \rangle^3 \over
\langle V_1, V_2 \rangle
\langle V_3, V_4 \rangle
\langle V_4, V_1 \rangle } =
 {[n,1]^3 \over \phi_1^3 \phi_n^3} . \label{eqac}
 \end{eqnarray}

Eq.~(\ref{eqac}) was not proved by doing the summation  directly.
In fact we prove it by analyzing its pole terms and prove that all
the pole terms are vanishing. The finite value can be obtained by
choosing a special configuration of $\phi_i$ and doing the
summation directly. We will do this in the next section.

By using eq.~(\ref{eqac}) in eq.~(\ref{eqab}), we have
\begin{equation}
A_n(1+,2-,\cdots, (n-1)-, n+) = { [n,1]^3 \over \prod_{i=1}^{n-1} [
i, i-1] } .
\end{equation}
This is the known result for the googly amplitude \cite{Parkea,
Giele}. It is the complex conjugate of the MHV amplitude,
eq.~(\ref{eqone}) for Minkowski signature.

 \begin{figure}[ht]
    \epsfxsize=100mm%
    \hfill\epsfbox{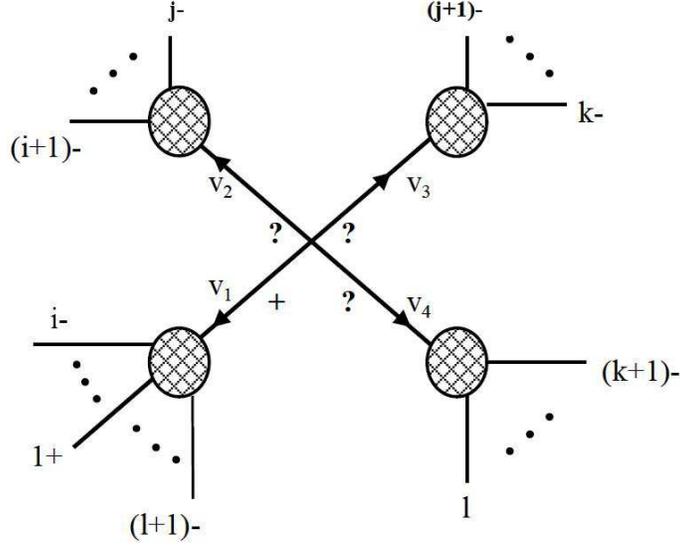}\hfill~\\
    \caption{The decomposition for the generic googly amplitude.
    We note that the helicity
     of the internal line can be different, depending where we
     put the positive helicity
     gluon. }
    \label{figa}
   \end{figure}

For the generic googly amplitude, the needed diagram decomposition
is given in Fig.~4. Depending on whether the positive helicity
gluon is in the second blob or in the third blob or in the last
blob, the expression for the 4 line vertex should be changed
appropriately. The 4-fold summation should give the generic googly
amplitude by using a more complicated identity. The required
identity is also proved by a similar analysis as given in the next
section  for the proof of eq~({\ref{eqac}). The details will be
given in a separate publication.

\section{The proof of eq.~(\ref{eqac})}

In this section we will prove eq.~(\ref{eqac}). As all  the
quantities are written in an $SL(2)$ invariant form, one can
choose a convenient $\tilde\eta$ to simplify the writing  of the
expressions. We will choose $\tilde\eta^1= 0 $ and $\tilde\eta^2
=1 $. Then we have
\begin{equation}
\phi_i = \tilde\lambda_{i2},
\end{equation}
and
\begin{equation}
{ {[}i,j] \over \phi_i \phi_j } =  {\tilde\lambda_{i1} \over
\tilde\lambda_{i2}} - {\tilde\lambda_{j1} \over
\tilde\lambda_{j2}}  .
\end{equation}

If we do a rescaling of $\tilde\lambda_{i1}$ by
$\tilde\lambda_{i2}$, i.e. by defining $\varphi_i =
{\tilde\lambda_{i1} \over \tilde\lambda_{i2}} $, and also do a
rescaling of $\lambda_{ia}$ by $1/\tilde\lambda_{i2}$, then
eq.~(\ref{eqac}) becomes:
\begin{eqnarray}
F(\varphi) & =  & \sum_{i=1}^{n-3}\sum_{j=i+1}^{n-2}\sum_{k=j+1}^{n-1}
(\varphi_i - \varphi_{i+1}) (\varphi_j - \varphi_{j+1})(\varphi_k -
\varphi_{k+1})
\, \nonumber \\
& & \qquad \times
{ \langle V_2, V_3 \rangle^3 \over
\langle V_1, V_2 \rangle
\langle V_3, V_4 \rangle
\langle V_4, V_1 \rangle } = (\varphi_n - \varphi_{1})^3 , \label{eqacc}
 \end{eqnarray}
where
\begin{eqnarray}
V_1  = \sum_{l=1}^i \lambda_l , \qquad & & V_2  = \sum_{l=i+1}^j \lambda_l ,
\\
V_3  = \sum_{l=j+1}^k \lambda_l , \qquad & & V_4  = \sum_{l=k+1}^n \lambda_l
.
\end{eqnarray}
There are also two constraints:
\begin{eqnarray}
 & & V_1 + V_2 + V_3 + V_4 = \sum_{l=1}^n \lambda_l = 0 , \label{eq29} \\
 & & \sum_{l= 1}^n \lambda_i\,  \varphi_i = 0 , \label{eq30}
 \end{eqnarray}
 from momentum conservation.

If we assume that $\lambda_1$ and $\lambda_n$ are solved
explicitly in terms of the rest $\lambda_{i}$ and all $\varphi_j$ (containing only
linear terms in $\varphi_{j}$ for $  2\le j\le n-1$), then the
left hand side of eq.~(\ref{eqacc}) can be considered as a
function of $\lambda_{j}$ for $  2\le j\le n-1$ and all
$\varphi_j$. As a function of $\varphi$ we will show that it is independent of
$\varphi_j$ for $  2\le j\le n-1$.

 \begin{figure}[ht]
    \epsfxsize=100mm%
    \hfill\epsfbox{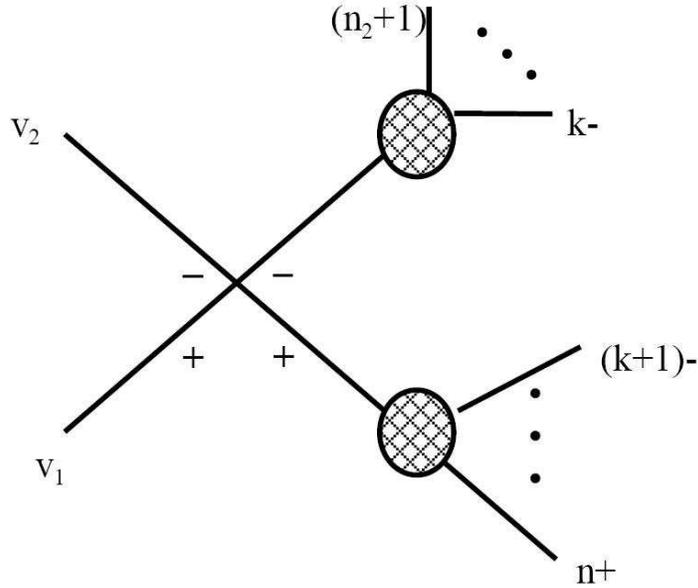}\hfill~\\
    \caption{The pole terms $\langle v_1,v_2\rangle$ from the factor $\langle V_1,
    V_2\rangle$. There is a summation over $k$.   }
    \label{figcc}
   \end{figure}

First one easily shows that there is no pole terms for $\varphi_j
\to \infty$ for $  2\le j\le n-1$. This is because there is at
most  a $\varphi_j^2$  in the numerator and the denominator factor
$\langle V_1, V_2 \rangle \langle V_3, V_4 \rangle \langle V_4,
V_1 \rangle $ grows as $\varphi_j^3$ because $V_1$ and $V_4$
contain linear terms in $\varphi_j$ and $V_2$ and $V_3$ don't
depend on $\varphi_j$. There is no $\varphi_j^2$ term in $\langle
V_4, V_1 \rangle $ although $V_1$ and $V_4$ both contain a
$\varphi_j \lambda_j$ term.

It remains to show all the finite pole terms are vanishing.
The possible pole terms appear if any factor of $\langle V_1, V_2 \rangle
\langle V_3, V_4 \rangle
\langle V_4, V_1 \rangle $ is vanishing. Let us consider first the vanishing of
$\langle V_1, V_2 \rangle$. We denote this set of $V_1$ and $V_2$ as $v_1$ and $v_2$:
\begin{equation}
v_1 = \lambda_1 + \cdots + \lambda_{n_1}, \qquad
v_2 = \lambda_{n_1+1} + \cdots + \lambda_{n_2}.
\end{equation}

As one can see from Fig.~5 that there are contributions from
summing over $k$ and fixing $i=n_1$ and $j=n_2$. The residues
(ignoring an overall factor $(\varphi_{n_1}- \varphi_{n_1+1})
(\varphi_{n_2}- \varphi_{n_2+1})$)  for these pole terms are:
\begin{equation}
C_1 =  \sum_{k=n_2 + 1}^{n-1} \,(\varphi_k - \varphi_{k+1})\,  { \langle  v_2,V_3\rangle^3
\over \langle v_2,V_3 \rangle \langle V_4,v_1 \rangle  } \, .
\end{equation}
Because $\langle v_1, v_2 \rangle = 0$ we have $v_1  = c v_2$. By
using this result  and the relation $v_1 + v_2 + V_3 + V_4 = 0$  in the above we have
\begin{eqnarray}
C_1 & = & {1\over c(c+1)} \sum_{k=n_2 + 1}^{n-1} \,(\varphi_k - \varphi_{k+1})\,
   \langle  v_2,V_3\rangle \nonumber \\
 & = &  {1\over c(c+1)}\left[ \sum_{k=n_2 + 1}^{n-1} \, \varphi_k \langle v_2,
 \lambda_ k \rangle - \phi_n \langle v_2,
 \lambda_{n_2+1} + \cdots + \lambda_{n-1} \rangle  \right] \nonumber \\
 & = & {1\over c(c+1)}  \sum_{k=n_2 + 1}^{n } \langle v_2,  \varphi_k \lambda_k
 \rangle .
 \end{eqnarray}

 \begin{figure}[ht]
    \epsfxsize=100mm%
    \hfill\epsfbox{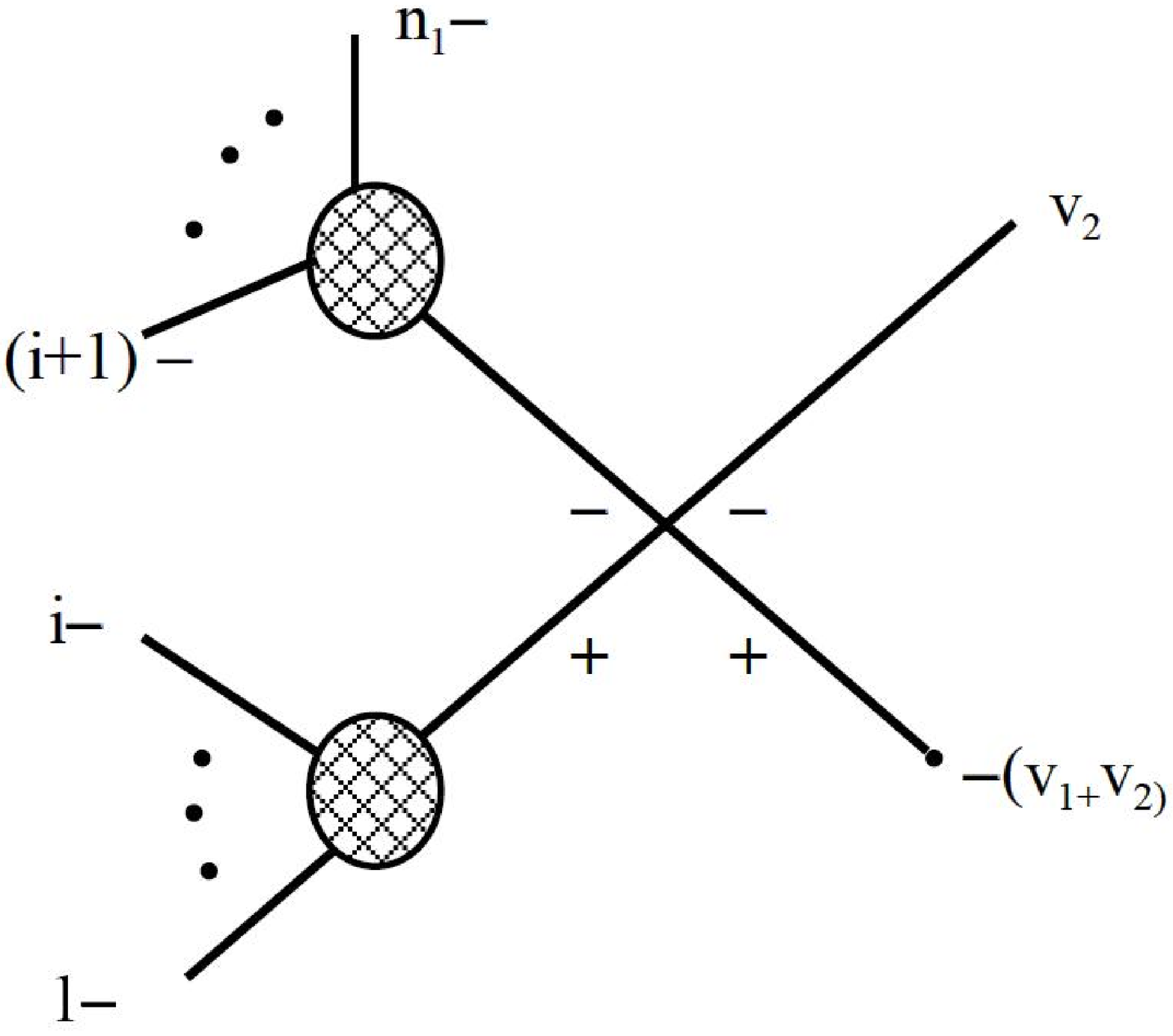}\hfill~\\
    \caption{The pole terms $\langle v_1,v_2\rangle$ from the factor $\langle V_3,
    V_4\rangle$. There is a summation over $i$.  }
    \label{figa}
   \end{figure}

Similar pole terms can also be obtained from the vanishing of the
factor $\langle V_3,V_4\rangle$ in Fig.~6 by setting $V_3 = v_2$
and $V_4 = -(v_1 + v_2)$ and summing over $i$. This give a
contribution:
\begin{eqnarray}
C_2 & = &  \sum_{i= 1}^{n_1-1} \,(\varphi_i - \varphi_{i+1})\,  { \langle  V_2,v_2\rangle^3
\over \langle V_1,V_2 \rangle \langle -(v_1+v_2),V_1 \rangle  } \,
\nonumber \\
& = & {1\over c(c+1)} \sum_{i= 1}^{n_1-1} \,(\varphi_i- \varphi_{i+1})\,
   \langle  v_2,V_1\rangle \nonumber \\
  & = & {1\over c(c+1)}  \sum_{i=  1}^{n_1 } \langle v_2,
  \varphi_i \lambda_i  \rangle .
 \end{eqnarray}

 \begin{figure}[ht]
    \epsfxsize=100mm%
    \hfill\epsfbox{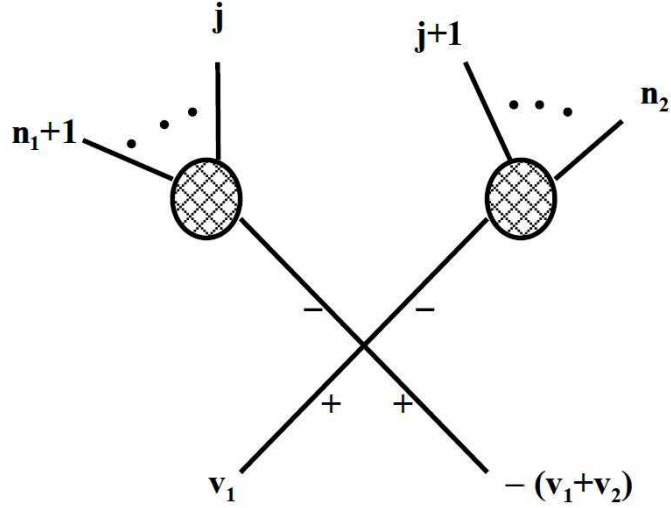}\hfill~\\
    \caption{The pole terms $\langle v_1,v_2\rangle$ from the factor $\langle V_4,
    V_1\rangle$. There is a summation over $j$.  }
    \label{figa}
   \end{figure}

The last piece of the pole terms is from the vanishing of the
factor $\langle V_4,V_1\rangle$ in Fig.~7 by setting $V_1 = v_1$
and $V_4 = -(v_1 + v_2)$  and summing over $j$. The result is:
\begin{equation}
C_3   =    {1\over c(c+1)}   \sum_{j= n_1 +1}^{n_2}    \langle v_2,
  \varphi_j \lambda_j  \rangle .
 \end{equation}

By summing the three residues together, we have
\begin{equation}
C_1 + C_2 + C_3 = {1\over c(c+1)}   \sum_{i=1}^n    \langle v_2,
  \varphi_i \lambda_i  \rangle = 0 ,
  \end{equation}
by using the constraint eq.~(\ref{eq30}). This proves that there
is no pole terms for the function $F(\phi)$. So $F(\phi)$ must be
independent of all $\varphi_j$ for $   2\le j\le n-1$.

Having proved the independence of $F(\varphi)$ on $\varphi_j$, let
us compute this function explicitly. As it is independent of
$\varphi_j$ for $  2\le j\le n-1$, we can choose a special set of
$\varphi $. A convenient choice is as follows:
\begin{equation}
\varphi_i = x, \qquad \hbox{and} \qquad \varphi_3 = \cdots =
\varphi_{n-1} = y.
\end{equation}
For generic $x$ and $y$ (and generic $\lambda_i$ which we don't
say explicityl), one can show that all the possible $\langle
V_1,V_2\rangle$, $\langle V_3,V_4\rangle$ and $\langle
V_4,V_1\rangle$ for different choices of $i$, $j$ and $k$ are
non-vanishing. For our choice of $\varphi$, the possible
non-vanishing factor for $(\varphi_i - \varphi_{i+1}) (\varphi_j -
\varphi_{j+1})(\varphi_k - \varphi_{k+1})$ is from  $i=1$, $j =2$
and $k=n-1$ only. We have then
\begin{equation}
F(\varphi)   =   (\varphi_1 - x)(x - y)(y-\varphi_n) \, { \langle
\lambda_2 , \lambda_3 + \cdots + \lambda_{n-1}  \rangle^3
\over \langle
\lambda_1 , \lambda_2     \rangle
\langle
  \lambda_3 + \cdots + \lambda_{n-1}, \lambda_n  \rangle
\langle
\lambda_n, \lambda_ 1   \rangle }  .
\end{equation}
By using the two constraints eq.~(\ref{eq29}) and
eq.~(\ref{eq30}), we have
\begin{equation}
\lambda_1 = { (x-\varphi_n)\lambda + (y-\varphi_n)\mu \over
\varphi_n - \varphi_1}, \qquad \lambda_n =  - {
(x-\varphi_1)\lambda + (y-\varphi_1)\mu \over \varphi_n -
\varphi_1}, \label{eq38}
\end{equation}
by setting $\lambda_2 = \lambda$ and $ \lambda_3 + \cdots +
\lambda_{n+1} = \mu$. By using these results in eq.~(\ref{eq38}),
we obtained exactly $F(\varphi) = (\varphi_n-\varphi_1)^3$. This
completes the proof of eq.~(\ref{eqac}).

\section*{Acknowledgments}

The author would like to thank Zhe Chang for encouragement and   Bin Chen and
Jun-Bao Wu for discussions. He would also like to thank Man-Lian
Zhang for drawing all the figures.

\end{document}